\documentclass[prl,twocolumn,amssymb,showpacs,superscriptaddress]{revtex4}
\usepackage{amsmath} 
\usepackage{amssymb} 
\usepackage{arydshln}
\usepackage{graphicx}
\usepackage{epsfig}
\usepackage{epstopdf}
\usepackage{color}
\usepackage{rotating}
\usepackage{bm}

\begin{document}
\title{Charge fluctuations in the unconventional metallic state of Li$_{0.9}$Mo$_6$O$_{17}$}
\author{J. Merino}
\affiliation{Departamento de F\'isica Te\'orica de la Materia Condensada, Condensed Matter Physics Center (IFIMAC)
and Instituto Nicol\'as Cabrera,
Universidad Aut\'onoma de Madrid, Madrid 28049, Spain}
\author{J.V. Alvarez} 
\affiliation{Departamento de F\'isica de la Materia Condensada, Condensed Matter Physics Center (IFIMAC)
and Instituto Nicol\'as Cabrera,
Universidad Aut\'onoma de Madrid, Madrid 28049, Spain}
\date{\today}
\begin{abstract}
Charge fluctuations in the quasi-one-dimensional material Li$_{0.9}$Mo$_6$O$_{17}$ are analyzed based on a multiorbital extended Hubbard model. A 
charge ordering transition induced by Coulomb repulsion is found with a charge ordering pattern different from a conventional charge density wave
driven by Fermi surface nesting. The metallic state displays a characteristic charge collective mode which softens signalling the proximity to the transition.
We argue that the strong scattering between electrons generated by these charge order fluctuations  
can lead to the unconventional metallic state observed above the superconducting transition temperature in Li$_{0.9}$Mo$_6$O$_{17}$. 
\end{abstract}
\maketitle

\paragraph{Introduction and motivation.}
Charge ordering phenomena is relevant to a wide range
of strongly correlated materials including copper-oxide (high-temperature) 
superconductors\cite{ghiringhelli2012,silvaneto2014,comin2014}, manganites\cite{cheong1998},  
sodium cobaltates \cite{cava2004} and the layered quarter-filled organic molecular 
conductors\cite{seo2006}.  In particular, for the cuprate superconductors charge ordering 
has been recently observed in the pseudogap region in close proximity to the superconducting 
phase raising questions about its relevance to the high-T$_c$ superconductivity\cite{kivelson2012}. 

The purple bronze Li$_{0.9}$Mo$_6$O$_{17}$  is a quasi-one-dimensional material
which displays behavior consistent with a Luttinger liquid \cite{wang2006,cazalilla2005} in a wide temperature range. When 
temperature is decreased an upturn of the resistivity occurs at $T_m \sim 20 $ K and the  material becomes 
superconducting at lower temperatures around $T_c \sim 1$ K \cite{schlenker1985}. The rather small enhancement 
of the resistivity below $T_m$  (just a factor of two) and the lack of spectral evidence of a gap makes 
it difficult to reconcile this upturn with an insulating phase. The fact that the resistivity is a 
decreasing function of temperature above the superconducting transition is in contrast with the superconducting transition
observed in conventional metals and in other strongly correlated materials except for the
quarter-filled organic materials\cite{mori2009,dressel2010}. Understanding the unconventional 
metallic state, whether it is a "bad" metal with incoherent excitations or not, may be crucial to the mechanism of 
superconductivity in Li$_{0.9}$Mo$_6$O$_{17}$. 

Conventional charge density waves (CDW) in solids involve a modulation of the electronic density accompanied by a crystal structure distortion. In Li$_{0.9}$Mo$_6$O$_{17}$ the resistivity upturn below $T_m$ is not accompanied by any structural transition as evidenced by high resolution X-ray scattering, neutron scattering,\cite{daluz2011} and thermal expansion\cite{dossantos2007} experiments. However, observing
a structural instability driven by Fermi surface nesting
requires a sufficiently large electron-lattice coupling which may not be present
in Li$_{0.9}$Mo$_6$O$_{17}$ as is also found in the family of quasi-one-dimensional TMTTF organic salts \cite{brown2000}.

In this Letter, we present a microscopic theory for the unconventional metallic properties observed in  
Li$_{0.9}$Mo$_6$O$_{17}$.  Based on a minimal extended Hubbard model recently 
introduced\cite{merino2012,chudzinsky2012} we show that Li$_{0.9}$Mo$_6$O$_{17}$ is close to a charge ordering (CO) transition
driven by the Coulomb repulsion. Using the random phase approximation (RPA), we identify the CO pattern characterized by 
the ordering wave vector, ${\bf Q}$, which is different from the conventional $2k_F$-CDW. The CO transition line, $T_{\rm CO}$,  displays 'reentrant' behavior which is responsible for a non-monotonic behavior of charge fluctuations being 
strongest around $T_m$ in the metallic phase close to CO. The dynamical charge
susceptibility displays a collective mode softening at momentum ${\bf Q}$ signalling the proximity to CO. 
We propose measurements of the dielectric constant \cite{monceau2000}, scanning tunneling microscopy (STM) and 
nuclear magnetic resonance (NMR) relaxation rate, 1/T$_1$T, to 
probe the $T$-dependence, strength and CO pattern of charge fluctuations in Li$_{0.9}$Mo$_6$O$_{17}$, testing
our predictions. The charge collective mode in the metallic phase can be explored with high resolution inelastic 
X-ray scattering (HRIX). 
In analogy with the effect of magnetic fluctuations in nearly antiferromagnetic metals, charge fluctuations can also 
lead to the unconventional $T$-dependence of the 
specific heat coefficient\cite{schlenker1985} and resistivity \cite{filippini1989,hussey2011} observed in  Li$_{0.9}$Mo$_6$O$_{17}$ 
assuming that the material is in the proximity but not neccesarily at CO. 
\begin{figure}
\epsfig{file=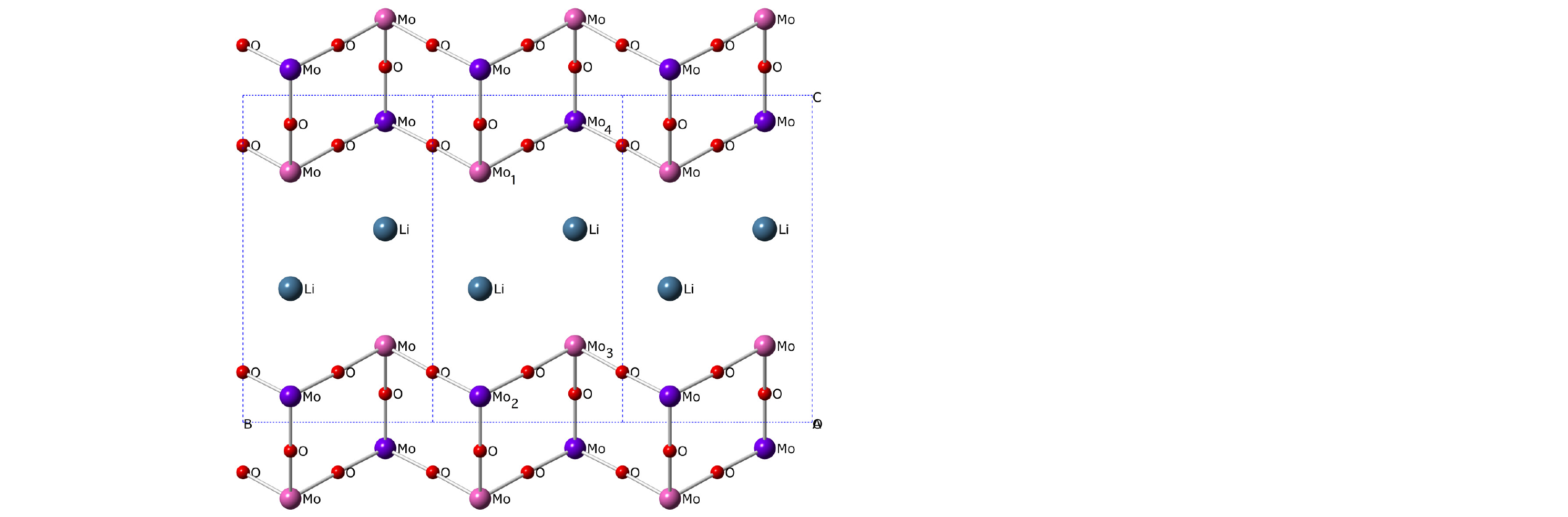,width=12cm,angle=0}
\caption{(Color online) Charge ordering phenomena in the extended Hubbard model (\ref{eq:model}) for Li$_{0.9}$Mo$_6$O$_{17}$.
We show the crystal structure of Li$_{0.9}$Mo$_6$O$_{17}$ projected onto the $b$-$c$ plane showing only the
Mo and O atoms forming the zig-zag ladders relevant to the low energy electronic properties.  The real space charge 
ordering pattern consisting of alternating charge-rich (purple) and charge-poor (magenta) Mo atoms arising in the model is also 
shown. 
 }
\label{fig:fig1}
\end{figure}

\begin{figure}
\epsfig{file=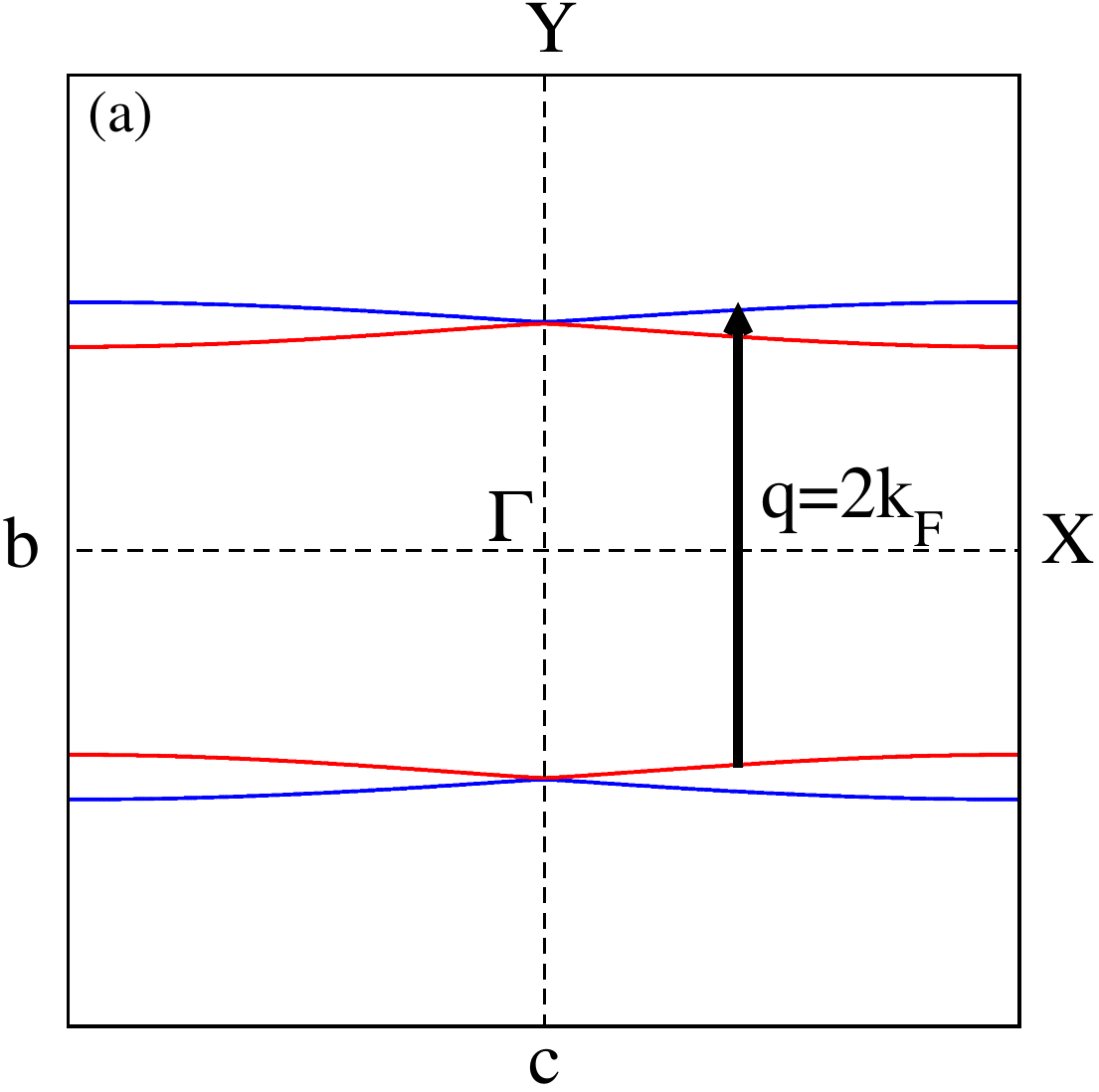,width=4.2cm,angle=0,clip=}
\epsfig{file=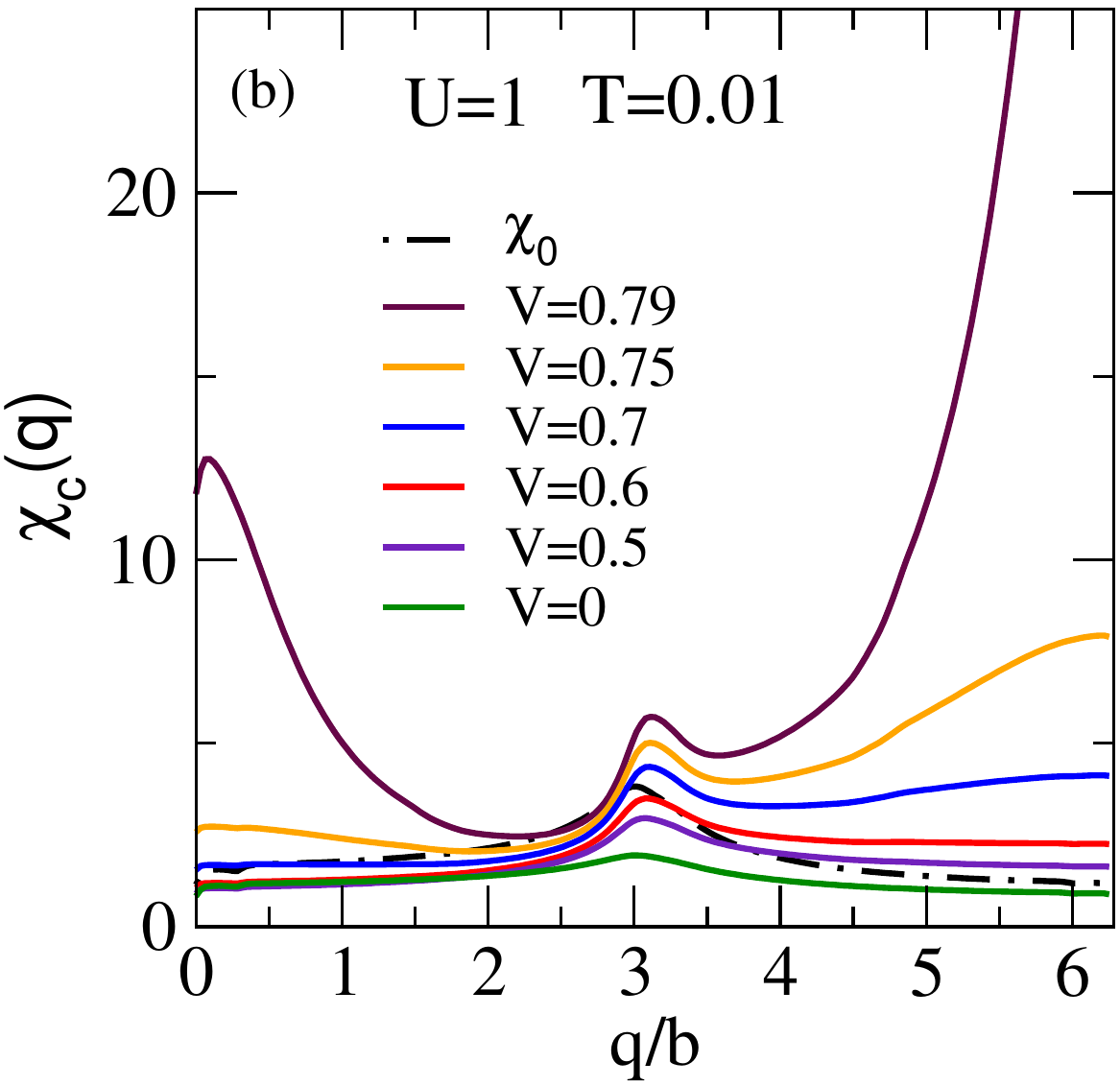,width=4.2cm,angle=0,clip=}
\caption{(Color online) Charge order instability induced by Coulomb repulsion. (a) The Fermi surface obtained from 
our effective model for Li$_{0.9}$Mo$_6$O$_{17}$ 
is shown. (b) The static charge susceptibility, $\chi_c({\bf q})$ along the $(0,\frac{q}{b},\frac{\pi}{c/2})$ direction
shows the rapid increase of charge fluctuations at wave vector: ${\bf Q}= (0,\frac{\pi}{b/2},\frac{\pi}{c/2})$ associated 
with the Coulomb induced CO occurring at $V_{\rm CO} \approx 0.8$. The smaller structure at ${\bf q}=(0,\frac{\pi}{b},\frac{\pi}{c/2})$ is related to Fermi surface nesting at $q=2k_F \lesssim  \pi/b$. }
\label{fig:fig2}
\end{figure}

\paragraph{Microscopic model.}
In Fig. \ref{fig:fig1} we show the zig-zag ladders consisting of Mo and O atoms which lead to the
characteristic quasi-one-dimensional electronic structure of the material. 
The minimal strongly correlated model which can capture the essential physics of Li$_{0.9}$Mo$_6$O$_{17}$  is a 
multi-orbital extended Hubbard model \cite{merino2012} which reads:
\begin{equation}
H=H_0+H_U,
\label{eq:model}
\end{equation}
where $H_0$ is the non-interacting tight-binding Hamiltonian: $H_0 = \sum_{i\alpha,j\beta} t_{i\alpha,j\beta} (d^{\dagger}_{i\alpha\sigma}d_{j\beta\sigma}+c.c.)$,
where  $d^{\dagger}_{i\alpha\sigma}$ creates an electron with spin $\sigma$ in a $d_{xy}$ orbital of a Mo atom, 
$\alpha$, which runs over four Mo atoms in the unit cell $i$ as shown in Fig. \ref{fig:fig1}. 
The one-electron hamiltonian can be expressed in terms of Bloch waves: 
$H_0=\sum_{{\bf k}\alpha \beta \sigma} t_{\alpha \beta} ({\bf k}) d^\dagger_{{\bf k} \alpha \sigma} d_{{\bf k} \beta \sigma}$ with the
following non-zero matrix elements\cite{merino2012}: $t_{12}({\bf k})=t_{34}({\bf k})=-t_\perp e^{i{\bf k}\cdot \delta_\perp},
t_{13}({\bf k})=-2t'e^{i{\bf k} \cdot {\bf \delta}_1}\cos({{\bf k}\cdot {\bf b} \over 2}),  t_{14}({\bf k})=-2tA(k_a,k_c)\cos({{\bf k}\cdot {\bf b} \over 2}), 
t_{23}({\bf k})=-2tA(k_a,k_c)\cos({{\bf k}\cdot {\bf b} \over 2})$, where $A(k_a,k_c)=e^{-i2\pi[0.1602k_a+0.1542k_c]}$, 
${\bf \delta}_\perp = 0.17{\bf a} -0.31{\bf c}$ and ${\bf \delta}_1=0.01{\bf a}+0.53{\bf c}$.
The momentum is expressed in terms of the unit cell coordinate system: ${\bf k} = k_a{\bf a} + k_b {\bf b} + k_c{\bf c}$,
and the hopping parameters are taken as: $t=0.5$ eV, between the nearest-neighbor Mo$_1$-Mo$_4$ and Mo$_2$-Mo$_3$ 
atoms along a chain, $t_\perp=-0.024$ eV between Mo$_1$-Mo$_2$ or Mo$_3$-Mo$_4$ atoms in a rung of a ladder
and $t'=0.036$ eV between Mo$_1$-Mo$_3$ atoms in neighboring zig-zag ladders (see Fig. \ref{fig:fig1}). 
The diagonalized hamiltonian: $H_0=\sum_{{\bf k}\mu \sigma} \epsilon_\mu({\bf k} ) d^\dagger_{{\bf k} \mu \sigma} d_{{\bf k} \mu \sigma}$, 
leads to four bands denoted by $\mu$, the two lowest ones crossing the Fermi energy\cite{popovic2006,merino2012}. The Fermi surface 
close to one quarter-filling, $n=0.45$, is shown in Fig. \ref{fig:fig2} (a).

The Coulomb interaction terms in the hamiltonian have been described previously in [\onlinecite{merino2012,chudzinsky2012}]
and read:
\begin{equation}
H_U=\sum_{ i\alpha,j\beta} U_{i\alpha j\beta}^{i\alpha j\beta}n_{i\alpha}n_{j\beta}.
\label{eq:hu}
\end{equation}
This term only includes the density-density Coulomb interaction contributions included in the extended Hubbard model. 
The Coulomb matrix elements in momentum space, $U_{\alpha\beta}^{\gamma\delta}({\bf q})=
U_{\alpha\beta}^{\alpha\beta}({\bf q}) \delta_{\alpha \gamma} \delta_{\beta \delta}$, 
have analogous expressions to the hopping terms but with the
diagonal Coulomb energies, $U_{\alpha\alpha}^{\alpha\alpha}({\bf q})=U/2$. Here, we
only consider the nearest-neighbor Coulomb interactions, $V=V_\perp$, within the zig-zag ladders
which leads to the CO pattern in Fig. \ref{fig:fig1}.   

\paragraph{Multiorbital RPA approach.}
The above model is analyzed based on a multi orbital random phase approximation (RPA) approach. 
The RPA charge susceptibility reads\cite{maier2009}: 
\begin{eqnarray}
&&(\chi_c)^{np}_{lm} ({\bf q},i\omega) =(\chi_0)^{np}_{lm}({\bf q},i\omega)
\nonumber \\
&&-\sum_{uvwz}(\chi_c)^{np}_{uv}({\bf q},i\omega)(U_c)^{uv}_{wz}({\bf q})(\chi_0)_{lm}^{wz}({\bf q},i\omega),
\end{eqnarray}
where the indices $l,m,n,p$ refer to the four Mo $d_{xy}$ orbitals present in the unit cell. 
$\hat{U}_c({\bf q})$ is the Coulomb matrix appearing in Eq. (\ref{eq:hu}) expressed in momentum space.  
The non-interacting susceptibility, $\chi_0$, reads: 

\begin{widetext}
\begin{equation}
(\chi_0)_{lm}^{wz}({\bf q}, i\omega)=-{2 \over N} \sum_{\bf k, \mu, \nu} {a^l_\mu({\bf k}) a^{w*}_\mu({\bf k})  a^{m}_\nu({\bf k + q}) a^{z*}_\nu({\bf k +q}) \over  i\omega +
\epsilon_\nu({\bf k +q}) -\epsilon_\mu({\bf k}) } [ f(\epsilon_\nu({\bf k+q}))-f(\epsilon_\mu({\bf k}) ) ] ,
\end{equation}
\end{widetext}
where $N$ is the number of lattice sites, $\nu,\mu$ are band indices.  The matrix elements $a^l_\mu({\bf k})=\langle l | \mu {\bf k}  \rangle $ are 
the coefficients of the eigenvectors diagonalizing $H_0$.

\paragraph{Charge ordering transition.}
In Fig. \ref{fig:fig2} we show the evolution of the static RPA charge susceptibility obtained from: $\chi_c({\bf q}) = 
\sum_{uw}(\chi_c)^{uu}_{ww}({\bf q},i 0^+)/2$,
with the offsite Coulomb repulsion $V$ and $U=1$. The susceptibility is evaluated along the $(0,\frac{q}{b},\frac{\pi}{c/2})$ direction in 
momentum space.  The charge susceptibility is 
enhanced with $V$ particularly at the wave vector ${\bf Q}= (0,\frac{\pi}{b/2},\frac{\pi}{c/2})$, which 
corresponds to having alternating charge rich and charge poor Mo atoms shown in Fig. \ref{fig:fig1}. This signals the charge 
ordering transition associated with the Coulomb repulsion (note that the unit cell defined by $a,b,c$ contains four Mo atoms). There 
is a smaller structure at  about $(0,\frac{\pi}{b},\frac{\pi}{c/2})$ associated 
with the nesting vector $2k_F$ connecting the different sections of the Fermi surface as shown in Fig. \ref{fig:fig2}. 

\begin{figure}
\epsfig{file=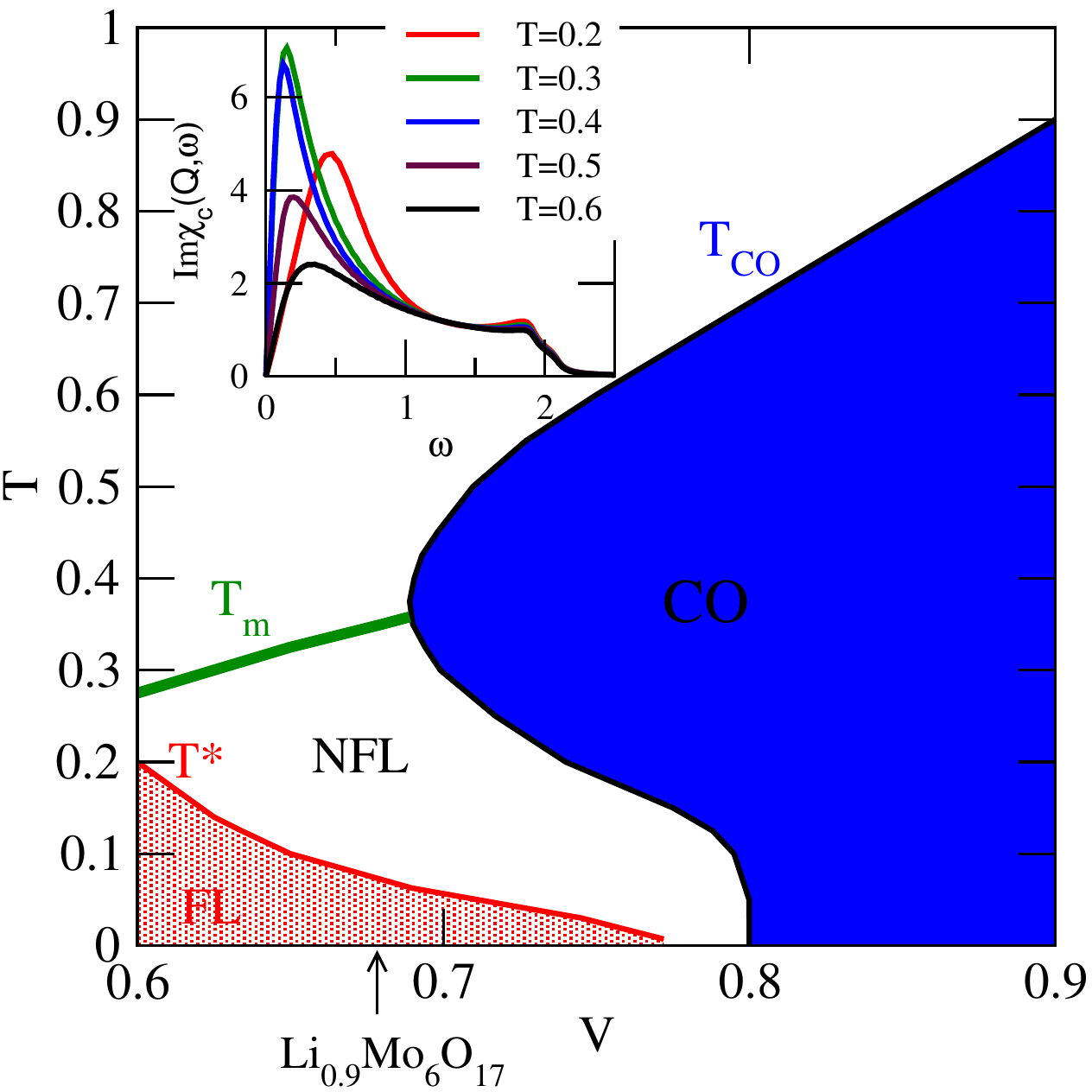,width=6cm,angle=0,clip=}
\caption{(Color online) Phase diagram of the effective extended Hubbard model for Li$_{0.9}$Mo$_6$O$_{17}$.  
The $T-V$ phase diagram obtained from RPA is shown for fixed value of $U=1$ and varying $V$. Charge ordered (CO) and
homogeneous metallic phases are separated by the CO transition line, $T_{\rm CO}$, which displays 'reentrant' behavior.
$T_m$ is the temperature scale associated with the onset of charge fluctuations. Fermi liquid (FL) and non 
Fermi liquid (NFL) phases are separated by the crossover scale $T^*$. The inset shows the $T$-dependence
of Im$\chi_c({\bf Q},\omega)$ in the metallic phase close to CO displaying the softening and enhancement 
of the charge collective mode around $T_m$. The proposed location for Li$_{0.9}$Mo$_6$O$_{17}$ at ambient 
pressure is marked with a vertical arrow. The system would effectively shift to smaller $V$  
under the effect of pressure.}
\label{fig:fig3}
\end{figure}

\begin{figure}
\epsfig{file=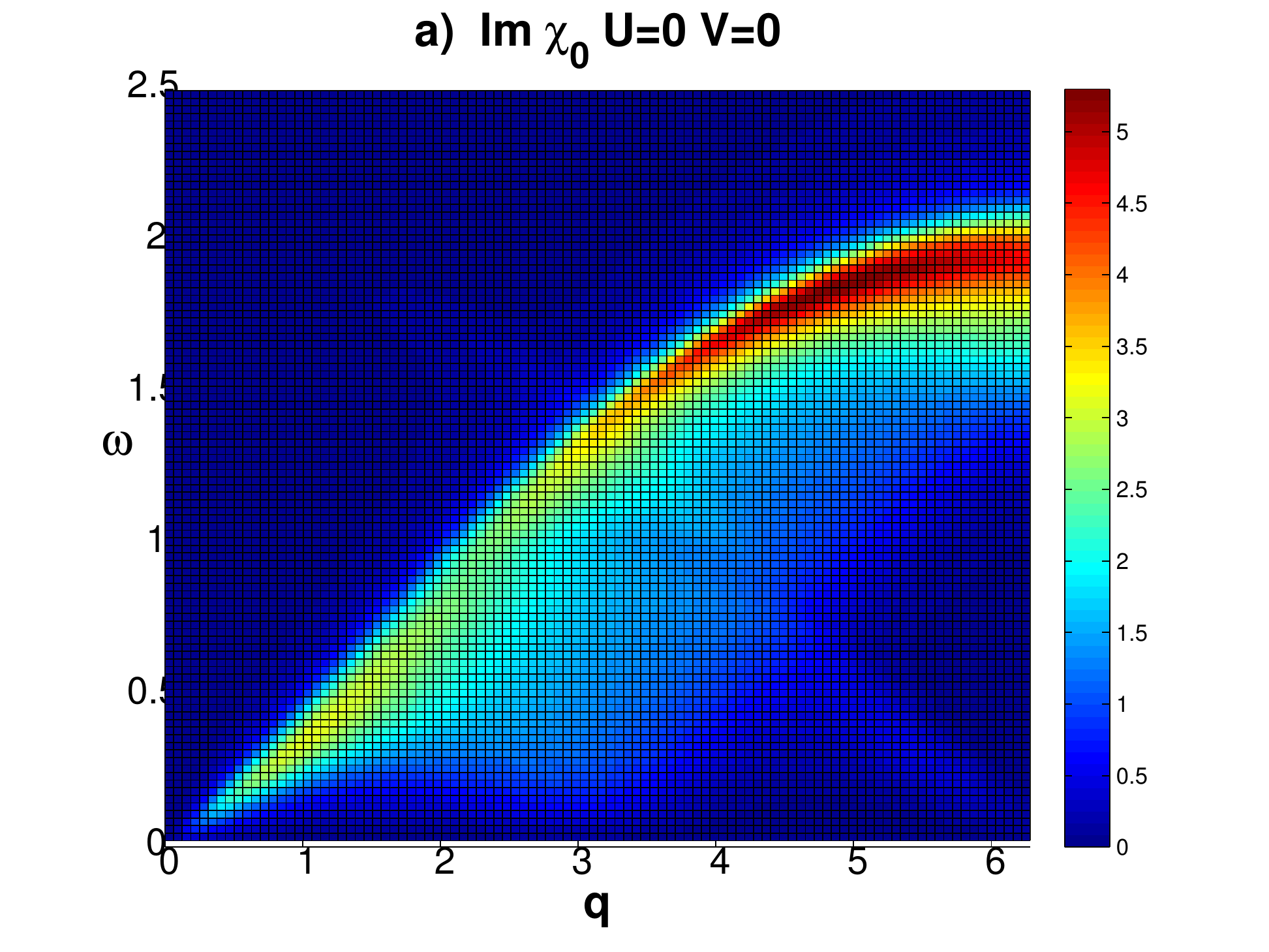,width=4.3cm,angle=0,clip=}
\epsfig{file=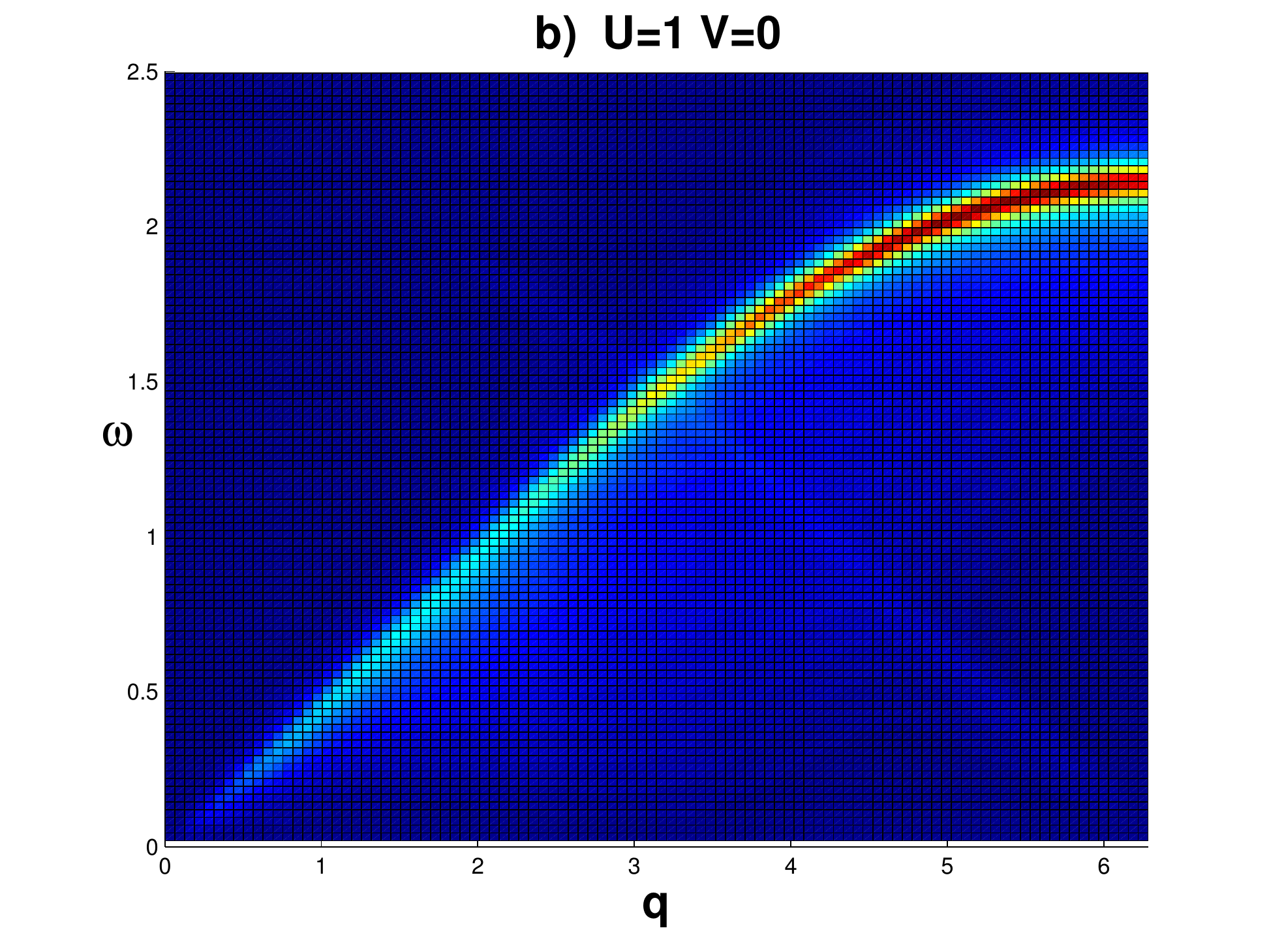,width=4.2cm,angle=0,clip=}
\epsfig{file=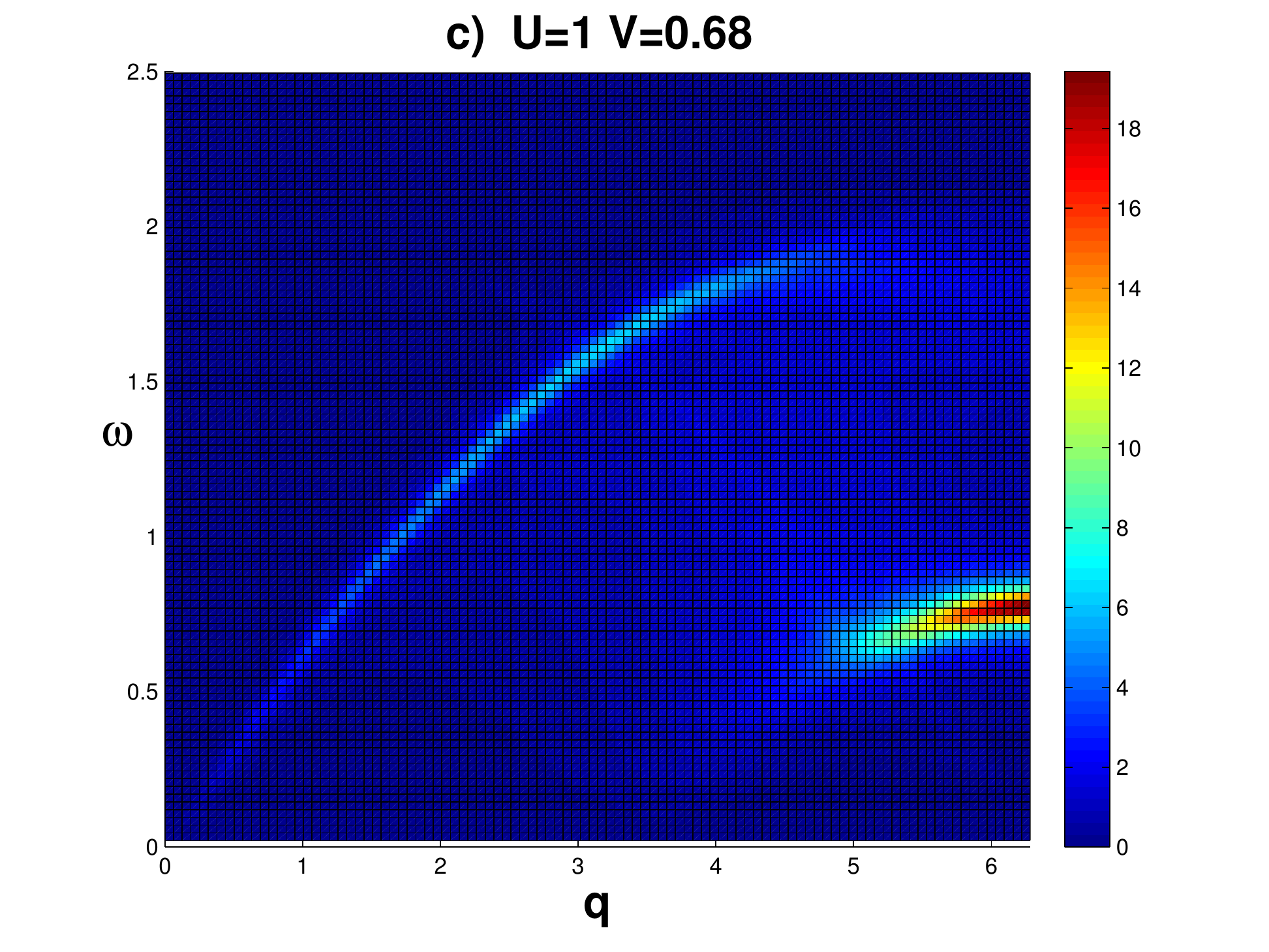,width=4.3cm,angle=0,clip=}
\epsfig{file=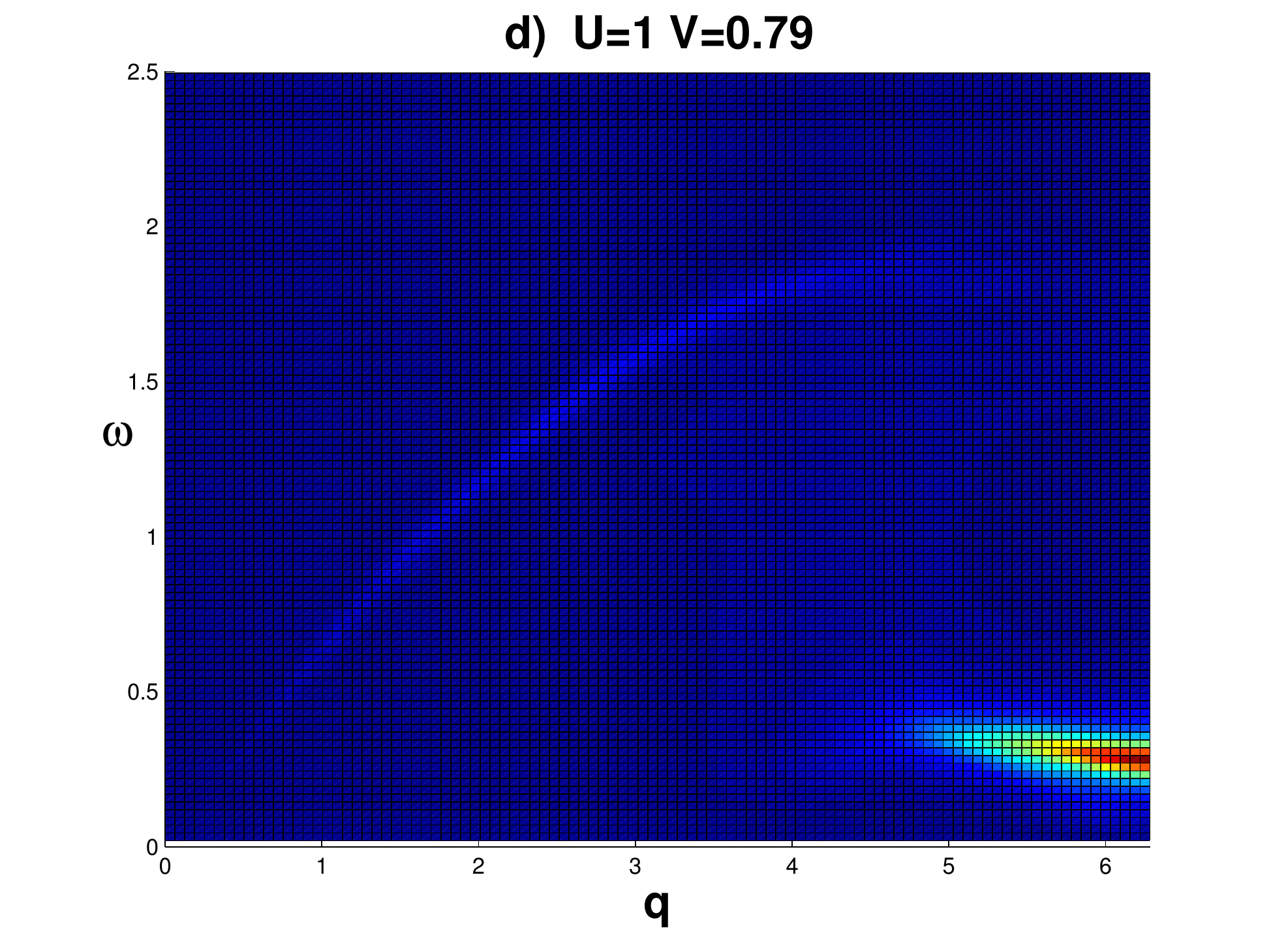,width=4.2cm,angle=0,clip=}

\caption{(Color online)  Imaginary part of the charge susceptibility Im$ \chi_c(q,\omega)$ showing the emergence and softening of the collective excitation as the interaction increases. a) Noninteracting charge susceptibility Im$ \chi_0(q,\omega)$ displaying the particle-hole continuum b) a Hubbard-like interaction ($U=1, V=0$)  c) an interaction compatible with purple bronze phenomenology $U=1, V=0.68$  and d) close to the CO transition $U=1, V=0.79$. Temperature is $T=0.05$.  }
\label{fig:fig4}
\end{figure}

\paragraph{Phase diagram.}
The $T-V$ phase diagram obtained from the present RPA approach is shown in Fig. \ref{fig:fig3}. The metallic
and charge ordered (CO) phases shown in Fig. \ref{fig:fig1} are separated by a 'reentrant' charge 
ordering transition line $T_{\rm CO}$. 
In the metallic phase, close to CO, charge fluctuations with the ${\bf Q}$ wave vector are strongly enhanced 
as shown in the inset of Fig. \ref{fig:fig4} for $V=0.68$ around our proposed location of Li$_{0.9}$Mo$_6$O$_{17}$ 
in the phase diagram. An enhancement of the low energy spectral 
weight occurs leading to a charge collective mode which softens 
and increases in amplitude as $T$ is decreased from about $T=0.6$. Such $T$-dependence 
of the charge collective mode occurs down to $T_m$, at which this behavior is reversed, {\it i. e.} the 
collective mode hardens
following the 'reentrant' shape of the transition line $T_{CO}$. Thus, $T_m$ is 
the temperature scale at which charge fluctuations induced by Coulomb repulsion 
are strongest in the homogeneous metallic phase sufficently close to CO. 





Close to the charge ordering instability, the electronic scattering by dynamical 
charge fluctuations\cite{castellani1995} largely influence the normal metallic properties.    
In analogy with nearly antiferromagnetic two-dimensional metals the scattering rate is non Fermi liquid,
$1/\tau(T)\propto T$,  with a crossover \cite{merino2006} to Fermi liquid behavior $1/\tau(T)\propto T^2$ below the
temperature $T^*$ (as obtained in Ref. [\onlinecite{merino2006}]). The temperature $T^*$ drops to zero close to the transition.
Our results are compatible with Luttinger liquid physics at higher temperatures although describing 
the crossover from one NFL to the other requires more sophisticated theoretical tools.  We also expect that the particular $T$-dependence of the charge fluctuations, leads to resistivity and specific heat slope \cite{schlenker1985} enhancements
around $T_m$. Furthermore, we find that the phase diagram
in Fig. \ref{fig:fig3} qualitatively agrees with the key features of Li$_{0.9}$Mo$_6$O$_{17}$. Under an external pressure, 
the temperature for the resistivity upturn, $T_m$, is suppressed \cite{filippini1989}, whereas the crossover 
temperature, $T^*$, for Fermi liquid behavior increases. The observed behavior of $T_m$ and $T^*$ with pressure is consistent with the phase diagram shown in Fig. \ref{fig:fig3}, since applying pressure is equivalent to reducing $V$ in our model.
Electronic localization effects arising close to CO not included in the present approach shoul lead to strong 
suppression of transition temperatures and $T_m$.

\paragraph{Softening of charge collective mode close to CO. }

In Fig. \ref{fig:fig4} we analyze Im$\chi_c({\bf q},\omega)$ as 
the system is driven close to CO at low temperature $T=0.05$. For non-interacting electrons, $U=V=0$, spectral weight is 
found in the particle-hole continuum with small weight in the region between 0 and 2$k_F$ as
expected for quasi-one-dimensional systems. Once the on-site Coulomb repulsion is turned on
spectral weight is enhanced around the highest energy branch of the particle-hole continuum due
to particle-hole excitations promoted by $U$. As $V$ is increased 
a redistribution of charge spectra around ${\bf Q}$ occurs in which particle-hole 
excitations of lower and lower energies gain weight. At $V=0.68$ 
a collective charge fluctuation mode is clearly
Close to the CO transition, $V\lesssim V_{\rm CO}$, the collective mode amplitude increases
shifting to zero energy signalling the Coulomb driven CO transition.
The behavior found for Im$\chi_c({\bf q}, \omega)$ close to CO can be understood from 
the singular part of the charge susceptibility in a two-dimensional system: $\chi_c({\bf q}, \omega) 
\approx {A \over  i \omega -\omega_{\bf q}}$,  
with $\omega_{\bf q}=\omega_0+C|{\bf q}-{\bf Q}|^2$, where $\omega_0 \rightarrow 0$ as 
the CO boundary is approached\cite{castellani1995} and $C$ a positive constant. 

The dynamical charge response of the system, Im$\chi_c({\bf q},\omega)$, can be experimentally analyzed 
using HRIX.  In particular, the dispersion of the collective mode discussed above
around ${\bf Q}$ could be extracted probing the proximity of the system to CO. 
Analogous plasmon softening around $2k_F$ has been observed with inelastic electron scattering on materials 
driven close to a conventional charge density wave \cite{jasper2011} instability. 
Measurements of the real part of the dielectric constant of the material could also track the enhancement 
in charge fluctuations occurring around $T_m$ as has been done in quasi-one-dimensional organic systems.\cite{monceau2000}.

\paragraph{Concluding remarks.}

We propose a new framework to describe the anomalous metallic behavior of the quasi-one-dimensional Li$_{0.9}$Mo$_6$O$_{17}$ at the temperatures $T \approx T_m$ around the  resistivity upturn. The small resistivity enhancement
acompanied with a weak feature on the specific heat are not consistent with an insulating state. There are neither signatures of a structural 
distortion nor convincing evidence of another phase transition. Our analysis shows that these anomalies can be attributed to strong charge 
fluctuations associated with CO induced by Coulomb repulsion and manifest themselves through a low energy collective excitation. Experimental 
probes such as NMR-1/T$_1$T relaxation rate, HRIX, STM and measurements of the dielectric constant can be used to search for 
fingerprints of such CO fluctuations.

\section*{Acknowledgments.} 
We acknowledge financial support from MINECO, MAT2012-37263-C02-01 (J.M.) and FIS2012-37549-C05-03 (J.V.A.).

\end{document}